\documentstyle[aps,prl,psfig]{revtex}

\def \missing  {$E_{T} \mbox{\hspace{-0.43cm}}/ \mbox{\hspace{0.4cm}}$}

\begin{document}

\title{
   Search for Second and Third Generation Leptoquarks
   Including Production via Technicolor Interactions
   in $p \bar{p}$ collisions at $\sqrt{s}=1.8$ TeV
}

\maketitle


\font\eightit=cmti8
\def\r#1{\ignorespaces $^{#1}$}
\hfilneg
\begin{sloppypar}
\noindent
T.~Affolder,\r {21} H.~Akimoto,\r {43}
A.~Akopian,\r {36} M.~G.~Albrow,\r {10} P.~Amaral,\r 7 S.~R.~Amendolia,\r {32} 
D.~Amidei,\r {24} K.~Anikeev,\r {22} J.~Antos,\r 1 
G.~Apollinari,\r {10}
% ...
\end{sloppypar}
\vskip .026in
\begin{center}
(CDF Collaboration)
\end{center}

\vskip .026in
\begin{center}
\r 1  {\eightit Institute of Physics, Academia Sinica, Taipei, Taiwan 11529, 
Republic of China} \\
\r 2  {\eightit Argonne National Laboratory, Argonne, Illinois 60439} \\
\r 3  {\eightit Istituto Nazionale di Fisica Nucleare, University of Bologna,
I-40127 Bologna, Italy} \\
\r 4  {\eightit Brandeis University, Waltham, Massachusetts 02254} \\
\r 5  {\eightit University of California at Los Angeles, Los 
Angeles, California  90024} \\  
\r 6  {\eightit Instituto de Fisica de Cantabria, University of Cantabria, 
39005 Santander, Spain} \\
\r 7  {\eightit Enrico Fermi Institute, University of Chicago, Chicago, 
Illinois 60637} \\
\r 8  {\eightit Joint Institute for Nuclear Research, RU-141980 Dubna, Russia}
\\
\r 9  {\eightit Duke University, Durham, North Carolina  27708} \\
\r {10}  {\eightit Fermi National Accelerator Laboratory, Batavia, Illinois 
60510} \\
\r {11} {\eightit University of Florida, Gainesville, Florida  32611} \\
\r {12} {\eightit Laboratori Nazionali di Frascati, Istituto Nazionale di Fisica
               Nucleare, I-00044 Frascati, Italy} \\
\r {13} {\eightit University of Geneva, CH-1211 Geneva 4, Switzerland} \\
\r {14} {\eightit Harvard University, Cambridge, Massachusetts 02138} \\
\r {15} {\eightit Hiroshima University, Higashi-Hiroshima 724, Japan} \\
\r {16} {\eightit University of Illinois, Urbana, Illinois 61801} \\
\r {17} {\eightit The Johns Hopkins University, Baltimore, Maryland 21218} \\
\r {18} {\eightit Institut f\"{u}r Experimentelle Kernphysik, 
Universit\"{a}t Karlsruhe, 76128 Karlsruhe, Germany} \\
\r {19} {\eightit Korean Hadron Collider Laboratory: Kyungpook National
University, Taegu 702-701; Seoul National University, Seoul 151-742; and
SungKyunKwan University, Suwon 440-746; Korea} \\
\r {20} {\eightit High Energy Accelerator Research Organization (KEK), Tsukuba, 
Ibaraki 305, Japan} \\
\r {21} {\eightit Ernest Orlando Lawrence Berkeley National Laboratory, 
Berkeley, California 94720} \\
\r {22} {\eightit Massachusetts Institute of Technology, Cambridge,
Massachusetts  02139} \\   
\r {23} {\eightit Institute of Particle Physics: McGill University, Montreal 
H3A 2T8; and University of Toronto, Toronto M5S 1A7; Canada} \\
\r {24} {\eightit University of Michigan, Ann Arbor, Michigan 48109} \\
\r {25} {\eightit Michigan State University, East Lansing, Michigan  48824} \\
\r {26} {\eightit University of New Mexico, Albuquerque, New Mexico 87131} \\
\r {27} {\eightit The Ohio State University, Columbus, Ohio  43210} \\
\r {28} {\eightit Osaka City University, Osaka 588, Japan} \\
\r {29} {\eightit University of Oxford, Oxford OX1 3RH, United Kingdom} \\
\r {30} {\eightit Universita di Padova, Istituto Nazionale di Fisica 
          Nucleare, Sezione di Padova, I-35131 Padova, Italy} \\
\r {31} {\eightit University of Pennsylvania, Philadelphia, 
        Pennsylvania 19104} \\   
\r {32} {\eightit Istituto Nazionale di Fisica Nucleare, University and Scuola
               Normale Superiore of Pisa, I-56100 Pisa, Italy} \\
\r {33} {\eightit University of Pittsburgh, Pittsburgh, Pennsylvania 15260} \\
\r {34} {\eightit Purdue University, West Lafayette, Indiana 47907} \\
\r {35} {\eightit University of Rochester, Rochester, New York 14627} \\
\r {36} {\eightit Rockefeller University, New York, New York 10021} \\
\r {37} {\eightit Rutgers University, Piscataway, New Jersey 08855} \\
\r {38} {\eightit Texas A\&M University, College Station, Texas 77843} \\
\r {39} {\eightit Texas Tech University, Lubbock, Texas 79409} \\
\r {40} {\eightit Istituto Nazionale di Fisica Nucleare, University of Trieste/
Udine, Italy} \\
\r {41} {\eightit University of Tsukuba, Tsukuba, Ibaraki 305, Japan} \\
\r {42} {\eightit Tufts University, Medford, Massachusetts 02155} \\
\r {43} {\eightit Waseda University, Tokyo 169, Japan} \\
\r {44} {\eightit University of Wisconsin, Madison, Wisconsin 53706} \\
\r {45} {\eightit Yale University, New Haven, Connecticut 06520} \\
\end{center}

\vspace{1.cm}

\begin{abstract}
\baselineskip 16pt


We report the results of a search for second and third generation
leptoquarks using 88 $\mbox{pb}^{-1}$ of data recorded by the Collider
Detector at Fermilab. Color triplet technipions, which play
the role of scalar leptoquarks, are investigated due to their 
potential production in decays of strongly coupled color octet 
technirhos. Events with a signature of two heavy flavor jets and 
missing energy may indicate the decay of a second (third) generation 
leptoquark to a charm (bottom) quark and a neutrino.
As the data is found to be consistent with Standard Model
expectations, mass limits are determined.


\vspace{0.5cm}


\end{abstract}

\baselineskip 16pt

\vspace{0.3cm}


While limited to interactions via gauge bosons in the Standard Model, 
quarks and leptons couple directly in theories with leptoquarks
\cite{LQs,preons,TC,farhisus,lane_ramana,genLQ}. 
The leptoquarks found in various models generally share similar
characteristics. 
They appear as color triplet bosons allowing for a Yukawa coupling of
strength $\lambda$ between quarks and leptons. In order to avoid the
constraints of proton decay, the interactions are assumed to conserve
baryon and lepton number. In addition, leptoquarks are typically
assumed to couple to fermions of the same generation in order to suppress
flavor changing neutral currents (FCNCs) \cite{davidson}. 
The principal mechanisms for leptoquark pair production at the Tevatron are 
$q \bar{q}$ annihilation and gluon fusion 
through either direct coupling to the gluon (``continuum'') 
or a technicolor resonance state.


The characteristics of leptoquark production from continuum can 
be categorized according to spin.
For scalar leptoquarks, the production cross section is parameter 
free since the coupling between leptoquark and gluon is determined by the
gauge symmetries of quantum chromodynamics (QCD) \cite{hew}.
Those interactions involving quark-lepton-leptoquark vertices 
are neglected due to the assumed small relative value of $\lambda$.
The scalar leptoquark production cross section is known to 
next-to-leading order \cite{kraemer}. 
Vector leptoquark interactions include
anomalous couplings to the gluons denoted as $\kappa_{G}$ and
$\lambda_{G}$ which are related to the anomalous `magnetic' 
moment and the `electric' quadrupole moment in the color field
\cite{blum}. $\kappa_{G}$ and $\lambda_{G}$ are associated with
different gauge fields and are considered independent in order to reduce
the reliance on any one specific model.
Probing the conspicuous choices leads to the investigation of
Yang-Mills type coupling when $\kappa_{G} = \lambda_{G} = 0$
and minimal coupling when $\kappa_{G} = 1$ and $\lambda_{G} = 0$.
At present only leading order processes have been calculated for
vector leptoquark pair production \cite{blum}.
The phenomenological parameter $\beta$ describes the branching
fraction of a leptoquark decaying to a final state which includes 
a charged lepton. 
Previous CDF analyses have examined $\beta=1$ and excluded leptoquark 
masses at the 95\% confidence level (CL) for the second
generation with scalar coupling of 202 GeV/c$^{2}$ and for the third
generation with scalar coupling of 99 GeV/c$^{2}$, minimal vector
coupling of 170 GeV/c$^{2}$, and Yang-Mills coupling of 
225 GeV/c$^{2}$ \cite{prevbeta1}.
For $\beta=0$, a search conducted by the 
D$\emptyset$ Collaboration 
has set limits at the 95\% CL
for second generation leptoquarks of 79, 160, and 205 GeV/c$^{2}$
\cite{lqdarin}
and for third generation leptoquarks of 94, 148, and 216 GeV/c$^{2}$
\cite{d0lq}
for scalar, minimal coupling, and Yang-Mills coupling, respectively.


Enhancement of leptoquark pair production occurs through the decay of
technicolor resonance states. 
Obviating the need for elementary scalar bosons, technicolor theories
present a dynamical explanation for electroweak symmetry breaking in
which quark and lepton chiral symmetries are explicitly broken by
gauge interactions including extended technicolor with a coupling
constant that evolves slowly to suppress FCNCs 
\cite{TC,farhisus,lane_ramana}.
In one of the established formulations, a complete family of
technifermions composed of an isodoublet of color triplet techniquarks
and an isodoublet of color singlet technileptons form a rich spectrum
of technimesons \cite{farhisus}. 
Combinations of techniquarks form color octet technirhos,
$\rho_{T8}$, some of which are endowed with the same quantum numbers
as the gluon allowing for $s$-channel coupling.
The color triplet and octet technipions, denoted by
$\pi_{LQ}$ and $\pi_{T8}$, couple in a 
Higgs-like fashion to quarks and leptons and thus are expected to 
decay into heavy fermion pairs. The $\pi_{LQ}$ is identified as a
scalar leptoquark. 
Contingent on phase space, the $\rho_{T8}$
may decay to quark, gluon, $\pi_{LQ}$, or $\pi_{T8}$ pairs. 
The leading-order cross section for leptoquark pair production
from technirho resonance depends upon the technirho and leptoquark
masses and the technirho width \cite{lane_ramana}. 
The technirho width is sensitive to changes in mass difference between
the color octet technipion and leptoquark, 
$\Delta M = M(\pi_{T8}) - M(\pi_{LQ})$. 
QCD corrections can be calculated to find an expected mass 
difference of 50 GeV/c$^{2}$ \cite{lane_ramana}.
For the restricted case $M(\rho_{T8}) < 2 M(\pi_{LQ})$, 
a previous CDF analysis of the dijet mass spectrum has already excluded
$260 < M(\rho_{T8}) < 480$ GeV/c$^{2}$ at the 95\% CL \cite{rhoprev}.
CDF has also conducted a search when decay to $\pi_{LQ}$ is
kinematically allowed for third generation leptoquark production 
using the $\tau \bar{\tau} b \bar{b}$ channel \cite{baum}.


The decay modes to $c \bar{\nu}$ and $b \bar{\nu}$
corresponding to $\beta = 0$
are utilized to search for pair produced leptoquarks in events with
two heavy flavor jets, missing transverse energy, and the absence of
high transverse momentum leptons. 
The continuum leptoquarks are assumed to be strictly second and
third generation, their decays involving $\nu_{\mu}$ and $\nu_{\tau}$ 
respectively.
Of the several potential color triplet technipion decays, the modes 
$\pi_{LQ} \rightarrow c \bar{\nu}_{\tau}$ for $M(\pi_{LQ})$ less than
the top quark mass and $\pi_{LQ} \rightarrow b \bar{\nu}_{\tau}$
are possible. Technically, the color triplet
technipion decaying to $c \bar{\nu}_{\tau}$ is a leptoquark of mixed 
generation. Yet since neutrino types cannot be distinguished in detector
events, these are considered to be similar to the second generation 
leptoquark. Questions concerning potential FCNC contributions arising
from this will not be addressed in the present work.


These signatures can be employed to conduct a 
search for leptoquark particles at CDF using a total integrated 
luminosity of $88.0 \pm 3.6 \; \mbox{pb}^{-1}$ collected during the 
1994-1995 Tevatron run. Since detailed descriptions of the CDF
detector and its components exist \cite{CDF}, only a
recapitulation follows. 
Detector positions are given by a coordinate system 
with the $z$ axis along the beamline, azimuthal angle, $\phi$, in the plane
transverse to the $z$ axis, and pseudorapidity, $\eta$.
Nearest to the interaction point, the silicon vertex detector 
(SVX$^{\prime}$) consists of four layers providing impact parameter
measurements with respect to the primary vertex in the plane
transverse to the beam direction \cite{svx}.
The primary vertices along the beam direction are reconstructed  
by the vertex tracking chamber in the region $|\eta | < 3.25$.
Directly inside a 1.4 T superconducting solenoidal magnet encompassing
a range $| \eta | < 1.1$ rests the central drift chamber used for
precision measurements of charged particles' transverse momenta. 
The calorimeter consists of electromagnetic 
and hadronic components covering a range $| \eta | < 4.2$.
The muon system covers a range of $| \eta | < 1$.
Missing transverse energy, \missing, indicating the presence of
neutrinos in the process, is the energy needed to balance 
the raw energy deposited in the calorimeter towers with $|\eta| < 3.6$
in the plane transverse to the beam direction.


Various selection criteria are applied to the data sample collected
using a trigger requiring \missing $> 35$ GeV.
Once the irrelevant sources of \missing originating from accelerator
induced and cosmic ray effects are removed, the remaining 304582
events are dominated by multijet QCD background.
Calorimeter information is used to determine jets through a fixed cone
algorithm \cite{cone} where the cone radius is 0.4 in $\eta - \phi$ space.
The raw energy deposited in the calorimeter towers is used to find the
jet energies. Events characterized by two or three hard jets with 
$E_{T} \geq 15$ GeV and $| \eta | \leq 2$ and no additional jets 
with $E_{T} > 7$ GeV and $| \eta | \leq 3.6$ are selected.
These requirements reduce both the background from $t \bar{t}$ events 
which typically results in four or more hard jets and the soft QCD
background arising from gluon radiation. 
To reduce systematic effects due to the trigger threshold, the
\missing trigger requirement is increased to  \missing $> 40$ GeV.
As jet energy mismeasurement results in missing energy appearing
parallel or anti-parallel to the jet direction, the
\missing direction is required to be well separated from any jet. 
To implement these criteria, the angles between \missing and any jet are
restricted to $\Delta \phi($\missing$, j) > 45^{\circ}$
and between \missing and the leading $E_{T}$ jet 
$\Delta \phi($\missing$,j_{1}) < 165^{\circ}$.
To further reduce QCD background, the angle between the two
highest $E_{T}$ jets is restricted to
$45^{\circ} < \Delta \phi(j_{1}, j_{2}) < 165^{\circ}$.
569 events pass these selections.
The $W$ and $Z$ backgrounds are lessened by rejecting events
containing loosely identified, high transverse momentum leptons
(excluding tau). Electron candidates are required to have lateral and
longitudinal shower profiles consistent with an electron \cite{top},
$E_{T} < 2 \cdot p_{T}$ when the momentum measurement is available,
and $E_{T} > 10$ GeV. 
Muon candidates are determined by matching a charged track to the 
calorimeter energy deposition compatible with a minimum ionizing 
particle \cite{top}.
If identified in the muon chambers, the muon candidates 
must have $p_{T} > 10$ GeV/c, otherwise they must have 
$p_{T} > 15$ GeV/c with additional $E_{T}$ as measured by the
calorimeter less than 5 GeV in a 0.4 radius cone around the lepton.
Once these criteria are employed, 396 events remain.

The technique of tagging $c$ and $b$ jets found in the recent CDF
scalar top and bottom quarks search \cite{stop} using the jet
probability algorithm \cite{jetprob} is employed.
The algorithm uses precision SVX$^{\prime}$ information to identify 
long-lived heavy quarks.
Jet probability, ${\cal P}_{jet}$, is derived from the combination 
of probabilities that
individual tracks come from a primary vertex (track probability)
for all tracks associated with a particular jet.
The algorithm constructs the probability that an ensemble of tracks in a jet
originates from a primary vertex.
For jets arising from primary vertices, ${\cal P}_{jet}$ is
flat from 0 to 1. When the jets emerge from secondary vertices, 
${\cal P}_{jet}$ peaks at 0. 
Events corresponding to leptoquark decays with 
a charm quark in the final state use the requirement of at least one 
taggable jet with ${\cal P}_{jet} \leq 0.05$. 
For the signatures with a bottom quark, at least one taggable jet with 
${\cal P}_{jet} \leq 0.01$ must be present. 
Applying these criteria, 11 observed events for the $c$ and 5 observed
events for the $b$ tagged data samples are found. 
The signal tagging efficiency is approximately 27\% for second 
generation leptoquarks and 49\% for third generation leptoquarks.


After the jet probability requirements are satisfied, the predominant 
background is determined to come from non-QCD sources. 
Events with $W$ and one jet, where the $W$ decays leptonically to a tau 
that decays hadronically, compose the largest single background with
$7.6$ and $3.0$ expected background events for ${\cal P}_{jet} \leq 0.05$
and $0.01$, respectively.
The total expected $W/Z/t \bar{t}/$diboson background for the $0.05$
jet probability cut is $11.1$ and for the $0.01$ cut is $4.5$. 
The QCD background comprises an expected $3.4$ events for 
${\cal P}_{jet} \leq 0.05$ and $1.3$ events for 
${\cal P}_{jet} \leq 0.01$.
Since the same data set and techniques are being employed as in
reference \cite{stop}, further discussion concerning backgrounds can
be found there.


Several Monte Carlo generators are employed together with a CDF
detector simulation package to both estimate the backgrounds and 
the expected signal.
The VECBOS program \cite{vecbos} allows for the tree-level calculation of
a vector boson plus jets production at the parton level.
The partons are then fragmented and hadronized using HERWIG 
routines \cite{herprt}.
Vector boson pair production and decay are simulated in ISAJET 
\cite{isajet}. HERWIG \cite{herprt} 
is employed to compute $t \bar{t}$ events.
To generate the signal events for scalar leptoquarks from continuum,
PYTHIA version 5.7 \cite{pythia} is used which already possesses the 
proper production cross section. For the vector leptoquarks, PYTHIA 
is encoded with the relevant cross section for leading order vector 
leptoquark production including anomalous couplings as derived 
in \cite{blum}. For the technicolor produced leptoquarks, 
the expected signal is generated by incorporating the cross 
sections and widths appropriate for color triplet technipion 
production \cite{lane_ramana} into PYTHIA. 
The parton distribution function employed in all these simulations 
was CTEQ 4L \cite{parton}. 
The renormalization scale for scalar leptoquark simulations was 
$\mu = M(LQ)$, for vector leptoquark simulations 
$\mu = \sqrt{\hat{s}}$, and for technicolor simulations 
$\mu = M(\pi_{LQ})$. 
The same search criteria are applied to the Monte Carlo samples as was
to the data.


Reflective of typical values, efficiencies for second generation 
leptoquarks produced from continuum are approximately 5\% 
for $M(LQ) = 125$ GeV/c$^{2}$, 
whereas in the case of production from technirho decay efficiencies
between 3 and 7\% are found, the efficiencies decreasing as the
technirho mass increases from 400 to 700 GeV/c$^{2}$.
Third generation leptoquarks with $M(LQ)=150$ GeV/c$^{2}$ are
found with approximately 10\% efficiency for continuum and 
as the technirho mass varies from 300 to 800 GeV/c$^{2}$, the
efficiency decreases from 12 to 3\%.
Since relative differences in efficiency between scalar and vector 
continuum produced leptoquarks are less than 5\% for 
$M(LQ) > 140$ GeV/c$^{2}$, the scalar leptoquark efficiency 
is employed for both.
The signal efficiency is degraded by a factor of $0.93$ to 
account for the effect of multiple $p \bar{p}$ interactions 
not present in the simulations \cite{stop}.


The systematic uncertainties for tagging efficiency, jet energy scale,
trigger, luminosity, and multiple interactions from \cite{stop} are
applicable to both the continuum and technicolor produced leptoquark
analyses and combine for an uncertainty of 18\%. Only those uncertainties
which are unique to leptoquarks will be mentioned below.
The dominant source of systematic uncertainty for continuum leptoquarks
comes from gluon radiation in the initial (ISR) and final (FSR) state. 
The systematic uncertainty was determined to be $31$\% by comparing 
efficiencies obtained with ISR or FSR neglected to those where ISR 
and FSR were included.
The effect of different choices of parton distribution function and
QCD renormalization scale is found to give a systematic uncertainty 
in efficiency of $10$\%. Combining these results with those of
\cite{stop}, the maximum total systematic uncertainty for the
efficiency is $37$\%.
For the leptoquarks generated from technirho decay, the systematic 
uncertainty due to ISR and FSR is found to be $25$\%.
The choice of parton distribution function and 
variation due to the renormalization scale set to 
$\mu = M(\pi_{LQ})/2$ and $2 M(\pi_{LQ})$
contribute a $9$\% and $20$\% systematic uncertainty,
respectively, to both the efficiencies and cross sections.
Consolidating the various results for the technicolor case, a
maximum total systematic uncertainty of $37$\% is likewise 
established.


For the second generation leptoquark search, 11 observed events 
are found and a background of $14.5 \pm 4.2$ events is estimated. 
In the case of third generation leptoquarks, 5 observed events 
are found and a background of $5.8 \pm 1.8$ events is estimated. 
As no excess of observed events over Standard Model background was
found, 95\% CL limits are determined through a background subtraction
method \cite{back} to ascertain excluded regions of parameter space.

\begin{figure}
   \centerline{\psfig{file=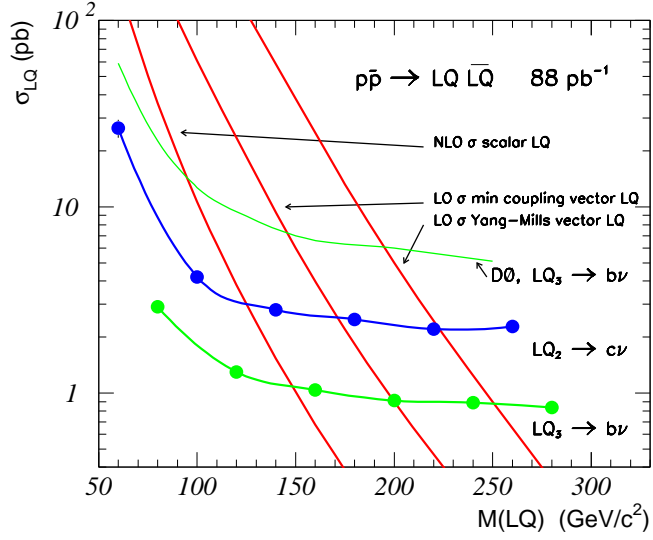,width=10cm}}
   \caption{
      95\% CL limit for scalar and vector second and third generation
      leptoquarks assuming $\beta=0$ compared to theoretical
      calculations. The 
      D$\emptyset$ Collaboration
      third generation leptoquark results are also shown [13].
   }
   \label{fig:CLcontin}

\end{figure}

For the continuum, a 95\% CL limit on scalar and vector leptoquark 
production cross sections for the various leptoquark types and masses
are determined. 
The results are shown in Figure \ref{fig:CLcontin}.
In the case of second generation leptoquarks, when these values are
compared to their corresponding theoretical cross sections
\cite{kraemer,blum} scalar leptoquarks with $M < 123$ GeV/c$^{2}$, 
minimally coupled vector leptoquarks with $M < 171$ GeV/c$^{2}$, and 
Yang-Mills vector leptoquarks with $M < 222$ GeV/c$^{2}$ are excluded.
Similarly, in the case of third generation leptoquarks,
scalar leptoquarks with $M < 148$ GeV/c$^{2}$,
minimally coupled vector leptoquarks with $M < 199$ GeV/c$^{2}$, and 
Yang-Mills vector leptoquarks with $M < 250$ GeV/c$^{2}$ are excluded.

\begin{figure}
   \centerline{\psfig{file=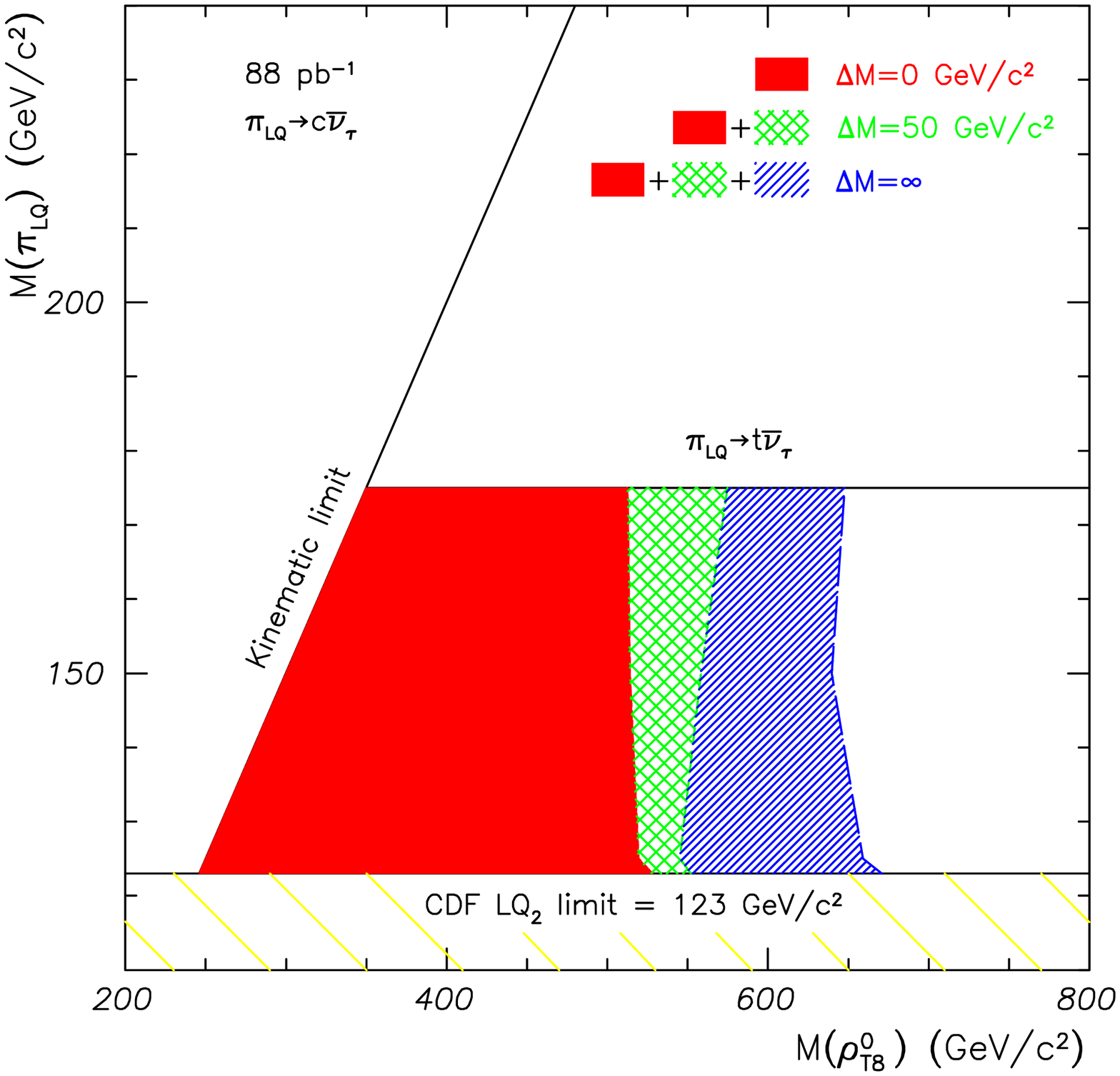,width=10cm}}
   \caption{
      95\% CL limit for the process
      $\rho_{T8}^{0} \rightarrow \pi_{LQ} \bar{\pi}_{LQ} \rightarrow
      c \bar{c} \nu_{\tau} \bar{\nu}_{\tau}$ at $\sqrt{s} = 1.8$ TeV. 
      The solid region corresponds to a mass difference of 
      $\Delta M = 0$ GeV/c$^{2}$, the solid and hatched regions 
      to $\Delta M = 50$ GeV/c$^{2}$, 
      and all three regions to $\Delta M = \infty$.
   }
   \label{fig:CLcnu}

\end{figure}

\begin{figure}
   \centerline{\psfig{file=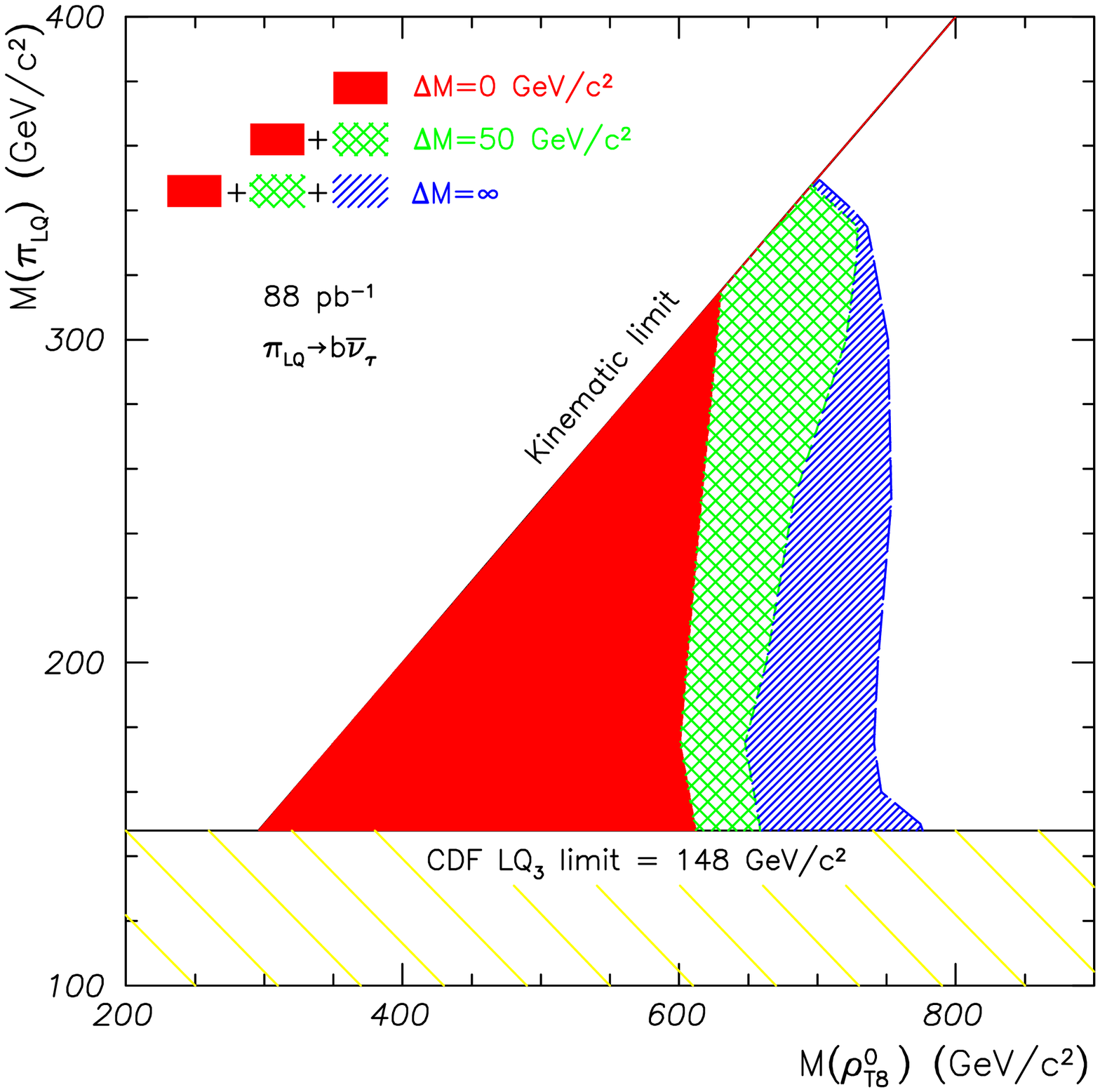,width=10cm}}

   \caption{
      95\% CL limit for the process
      $\rho_{T8}^{0} \rightarrow \pi_{LQ} \bar{\pi}_{LQ} \rightarrow
      b \bar{b} \nu_{\tau} \bar{\nu}_{\tau}$ at $\sqrt{s} = 1.8$ TeV. 
      The solid region corresponds to a mass difference of 
      $\Delta M = 0$ GeV/c$^{2}$, the solid and hatched regions 
      to $\Delta M = 50$ GeV/c$^{2}$, 
      and all three regions to $\Delta M = \infty$.
   }
   \label{fig:CL}

\end{figure}

The technicolor produced scalar leptoquarks entail a
complication in that the technirho mass is an additional parameter. 
Furthermore, since the production cross section is affected by
the mass difference $\Delta M$, the values $\Delta M = 0$, 
$50$ GeV/c$^{2}$, and $\infty$ are probed.
The 95\% CL exclusion regions in the $M(\rho_{T8}) - M(\pi_{LQ})$ plane 
shown as shaded areas in Figures \ref{fig:CLcnu} and \ref{fig:CL}
are determined by comparing the 95\% CL cross section limit for 
production of leptoquarks which decay to quarks and neutrinos 
to theoretical predictions \cite{lane_ramana}. 
The kinematically forbidden region is given by 
$M(\rho_{T8}) < 2 M (\pi_{LQ})$. The continuum scalar leptoquark 
limits found earlier in this analysis set an excluded mass for
second generation leptoquarks below 123 GeV/c$^{2}$ and for third generation
leptoquarks below 148 GeV/c$^{2}$. In Figure \ref{fig:CLcnu}, 
the decay of the leptoquark to
$c \bar{\nu}_{\tau}$ is limited by the top quark mass, above which the
leptoquark will decay preferentially to $t \bar{\nu}_{\tau}$. 
When $\Delta M = 0$, $M(\rho_{T8}) < 510$ GeV/c$^{2}$ for the second
generation and $M(\rho_{T8}) < 600$ GeV/c$^{2}$ for the third
generation are excluded at 95\% CL.

This analysis reports on the search for leptoquarks produced from
continuum and color octet technirho decays in $p \bar{p}$
collisions at $\sqrt{s} = 1.8$ TeV using 88 pb$^{-1}$ of data. 
Events with two or three jets, substantial missing energy, and no
high transverse momentum leptons are subjected to the jet probability 
requirement indicating at
least one jet being consistent with originating from a heavy flavor.
No excess of events above Standard Model predictions are found and
therefore 95\% CL limits are determined.


We thank the Fermilab staff and the technical staffs of the
participating institutions for their vital contributions. 
We also wish to thank Ken Lane for assistance in deriving technicolor
production cross sections.
This work was
supported by the U.S. Department of Energy and National Science Foundation;
the Italian Istituto Nazionale di Fisica Nucleare; the Ministry of Education,
Science, Sports and Culture of Japan; the Natural Sciences and Engineering
Research Council of Canada; the National Science Council of the Republic of
China; the Swiss National Science Foundation; the A. P. Sloan Foundation; the
Bundesministerium fuer Bildung und Forschung, Germany; and the Korea Science
and Engineering Foundation.



\newpage

\end{document}